\documentclass[11pt]{article}
\usepackage{amssymb}
\usepackage{amsmath}
\usepackage{amscd}
\usepackage{latexsym}

\catcode`@=11
\renewcommand{\section}{\@startsection{section}{1}{0pt}{\medskipamount}
{\medskipamount}{\large\bf}}
\numberwithin{equation}{section}
\catcode`@=12
\def\a{\alpha}

\def\g{\gamma}

\def\de{\delta}
\def\ve{\varepsilon}

\def\t{\theta}
\def\Th{\Theta}

\def\la{\lambda}

\def\s{\sigma}
\def\p{\phi}

\def\vk{\varkappa}

\def\ome{\omega}
\def\Om{\Omega}
\def\La{\Lambda}

\def\Ow{\widetilde \Omega}
\def\ot{\widetilde \omega}
\def\1{\bar 1}
\def\2{\bar 2}
\def\3{\bar 3}
\def\4{\bar 4}

\def\lrc{\,\lrcorner\,}
\def\hra{\,\hookrightarrow\,}

\newcommand{\ah}{\hat{a}}

\newcommand{\yb}{\bar{y}}
\newcommand{\ab}{\bar{\alpha}}
\newcommand{\bb}{\bar{b}}

\newcommand{\zb}{\bar{z}}
\newcommand{\zeb}{\bar{\zeta}}

\newcommand{\ib}{\bar{i}}
\newcommand{\jb}{\bar{j}}
\newcommand{\kb}{\bar{k}}

\newcommand{\C}{\mathbb C}
\newcommand{\R}{\mathbb R}

\newcommand{\Acal}{{\cal A}}
\newcommand{\Ncal}{{\cal N}}

\newcommand{\Fcal}{{\cal F}}
\newcommand{\Ecal}{{\cal E}}
\newcommand{\J}{{\cal J}}

\newcommand{\T}{{\cal T}}

\def\im{\mbox{i}}
\def\N2{$N{=}2$}
\def\pa{\mbox{$\partial$}}
\def\diff{\mbox{d}}
\def\tr{{\rm tr}}
\def\sfrac#1#2{{\textstyle\frac{#1}{#2}}}
\def\>{\rangle}
\def\<{\langle}
\def\+{\dagger}
\def\={\ =\ }
\def\und{\qquad\textrm{and}\qquad}
\def\and{\quad\textrm{and}\quad}
\def\for{\quad\textrm{for}\quad}

\begin{document}
\begin{titlepage}
\setcounter{page}{0}
\begin{flushright}
.
\end{flushright}

\vskip 2.5cm

\begin{center}

{\large\bf Non-Abelian Vortices, Super-Yang-Mills Theory and Spin(7)-Instantons
}

\vspace{12mm}
{\large Alexander D. Popov}
\\[2mm]
\noindent {\em Bogoliubov Laboratory of Theoretical Physics, JINR\\
141980 Dubna, Moscow Region, Russia}
\\
{Email: {\tt popov@theor.jinr.ru}}
\vspace{12mm}

\begin{abstract}
\noindent
We consider a complex vector bundle $\Ecal$ endowed with a connection
$\Acal$ over the eight-dimensional manifold $\R^2\times G/H$, where
$G/H=\,$SU(3)/U(1)$\times$U(1) is a homogeneous space provided with a
never integrable almost complex structure and a family of SU(3)-structures.
We establish an equivalence between $G$-invariant solutions $\Acal$ of the
Spin(7)-instanton equations on $\R^2\times G/H$ and general solutions of
non-Abelian coupled vortex equations on $\R^2$. These vortices are BPS
solitons in a $d=4$ gauge theory obtained from $\Ncal =1$ supersymmetric
Yang-Mills theory in ten dimensions compactified on the coset space $G/H$
with an SU(3)-structure. The novelty of the obtained vortex equations lies in
the fact that Higgs fields, defining morphisms of vector bundles over $\R^2$,
are not holomorphic in the generic case. Finally, we introduce BPS vortex equations
in $\Ncal =4$ super Yang-Mills theory and show that they have the same feature.

\end{abstract}
\end{center}

\end{titlepage}

\section{Introduction}

\noindent
Symmetries play an important role in physics and mathematics.
In particular, it is often useful to study solutions of partial
differential equations which are invariant under the action of
some symmetry group. Invariant solutions to the original equations
can be interpreted as ordinary solutions to a related set of
equations on the orbit space of the group action obtained with
the help of dimensional reduction.

In Yang-Mills theory in arbitrary dimensions, the $G$-equivariant
dimensional reduction on K\"ahler manifolds of the form
$M^{2n}{\times}G/H$, where $G/H$ is a K\"ahler homogeneous space
for a compact Lie group $G$ with a closed subgroup $H$, induces
a Yang-Mills-Higgs theory on $M^{2n}$ which is a quiver gauge theory
\cite{G1}-\cite{G5}. Recall that a quiver is an oriented graph, i.e.
a set of vertices $Q_0$ together with a set of arrows $Q_1$ between the
vertices. A path in $Q=(Q_0, Q_1)$ is a sequence of arrows in $Q_1$
which compose. A relation $r$ of the quiver is a formal finite sum of
paths and quivers with relations can naturally be associated with
any K\"ahler coset space $G/H$~\cite{G4, G5}. In a $G$-equivariant
dimensional reduction from $M^{2n}{\times}G/H$ to $M^{2n}$, one chooses
a quiver $Q=(Q_0, Q_1)$ associated with $G/H$ and then to each vertex from $Q_0$
there corresponds a vector bundle over $M^{2n}$ and to each arrow from $Q_1$
a morphism (defined by a Higgs field) of two bundles from the set $Q_0$.
The reduction of the associated first-order Hermitian-Yang-Mills equations
yields non-Abelian quiver vortex equations on $M^{2n}$~\cite{G1}-\cite{G5}.
The Seiberg-Witten monopole equations~\cite{Witten} for $n{=}2$ and the ordinary
vortex equations~\cite{Jaffe} for $n{=}1$ are particular instances of quiver
vortex equations. Recently, the formalism of $G$-equivariant reductions
has been applied in a variety of contexts~\cite{PS1}-\cite{DS}.

In this paper, we consider the manifold $X^8=\R^2\times\,$SU(3)/U(1)$\times$U(1),
and a complex vector bundle $\Ecal$ over $X^8$. However, instead of the
Hermitian-Yang-Mills equations on $\Ecal\to X^8$ which can be considered along
with the standard K\"ahler structures on $\R^2$ and SU(3)/U(1)$\times$U(1)
\cite{LPS4}, we consider the Spin(7)-instanton equations~\cite{DT, Lewis}, the
related SU(4)-instanton equations~\cite{DT} and their dimensional reduction to
$\R^2$. For that, on the coset space SU(3)/U(1)$\times$U(1), we introduce
a never integrable almost complex structure and a family of SU(3)-structures.
These SU(3)-structures induce a Spin(7)$\,\supset\,$SU(4) structure on the
direct product manifold $X^8=\R^2\times\,$SU(3)/U(1)$\times$U(1) with $\R^2\cong\C$.
Thus, our manifold $X^8$ is no longer K\"ahler and has a torsion. Using an
SU(4)-structure on $X^8$, we rewrite the Spin(7)-instanton equations for a
connection $\Acal$ on $\Ecal$ in the form of SU(4)-instanton equations.

The study of SU(3)-equivariant solutions to the SU(4)-instanton equations on
the manifold $\R^2\times\,$SU(3)/U(1)$\times$U(1) yields, via a dimensional reduction, a
new type of non-Abelian quiver vortex equations on $\R^2$. We write down these
vortex equations explicitly for a simplest quiver associated with the coset space
SU(3)/U(1)$\times$U(1).\footnote{A generalization to arbitrary quiver is
straightforward.} They are equations for connections $A^i$ on three vector bundles
$E_i$ over $\R^2$, $i=1,2,3$, linked by three homomorphisms $\p_i$ (Higgs fields)
which in general are not holomorphic. In fact, the Dolbeault operator $\bar\pa_A$
maps the Higgs fields $\p_i$ to polynomials of Higgs fields corresponding to
quiver relations $r_i$. This is a new potentiality of quiver vortex theory.
Holomorphicity is restored when a connection $\Acal$ on $\Ecal$ satisfies not
only the SU(4)-instanton equations but also the Hermitian-Yang-Mills equations
that in general is not possible.\footnote{There are SU(4)-instanton connections
which are not Hermitian-Yang-Mills.} Note that the considered quiver vortices are
BPS solitons in $\Ncal =1$ supersymmetric Yang-Mills theory in ten dimensions
reduced to $d=2$.

Finally, we consider the bosonic sector of $\Ncal =4$ U($k$) super Yang-Mills
(SYM) theory in Minkowski space $\R^{3,1}$ and introduce BPS vortex equations on
$\R^2\subset\R^{3,1}$. We show that these equations on a gauge potential $A$
on a U($k$)-bundle over $\R^2$ and three
complex matrix scalar fields $\p_i$ are also related to the SU(4)-instanton
equations in eight dimensions and have the same feature as the above quiver
vortex equations, i.e. the Higgs fields $\p_i$ are not holomorphic.

\bigskip

\section{SU(3)-structures on the homogeneous space SU(3)/U(1)$\times$U(1)}

\noindent
{\bf SU(3)-structure on 6-manifolds}.
Let $G$ be a closed subgroup of the orthogonal group SO($d$). A $G$-structure
on an oriented Riemannian manifold $(X^d, g)$ of dimension $d$ is a reduction
of the structure group SO($d$) of the tangent bundle of $X^d$ to the subgroup
$G$. In particular, for a six-dimensional manifold $(X^6, g)$, an SU(3)-structure
on $X^6$ is determined by a pair $(\ome , \Om )$, where $\ome$ is a non-degenerate
two-form (an almost symplectic structure) and $\Om$ is a decomposable complex
three-form such that
\begin{equation}\label{2.1}
\ome\wedge\Om =0\und\Om\wedge\bar\Om=-\frac{4\im}{3}\ \ome\wedge\ome\wedge\ome\ .
\end{equation}
Indeed, the above complex three-form
\begin{equation}\label{2.2}
\Om=\Th^1\wedge\Th^2\wedge\Th^3
\end{equation}
determines an almost complex structure $\J$ on $X^6$ such that
\begin{equation}\label{2.3}
\J\,\Th^i=\im\,\Th^i\for i=1,2,3\ ,
\end{equation}
i.e. forms $\Th^i$ span the space of forms of type (1,0). Hence,
$c_1(X^6)=0$ and the (3,0)-form $\Om$ is a global section of the
canonical bundle of $X^6$. For more details see e.g.~\cite{CS, But}
and references therein.

The form $\ome$ is of type (1,1) by virtue of (\ref{2.1}) and $g=\ome\J$
is an almost Hermitian metric. We choose
\begin{equation}\label{2.4}
g=\Th^1\Th^{\1}+\Th^2\Th^{\2}+\Th^3\Th^{\3}\and
\ome=\frac{\im}{2}\left (
\Th^1\wedge\Th^{\1}+\Th^2\wedge\Th^{\2}+\Th^3\wedge\Th^{\3}
\right )\ .
\end{equation}
We assume that $X^6$ is not a Calabi-Yau manifold and an almost
complex structure $\J$ is not integrable. Examples of such manifolds
are nearly K\"ahler~\cite{CS, But} and nearly Calabi-Yau~\cite{Xu} ones.

\smallskip

\noindent
{\bf Twistor space $\T(\C P^2)$}. We will consider a non-integrable almost complex
structure $\J$ and SU(3)-structures on the flag manifold SU(3)/U(1)$\times$U(1)
which is the twistor space $\T(\C P^2)$ of the projective plane $\C P^2$. It is fibred
over $\C P^2$ with the canonical projection
\begin{equation}\label{2.5}
\pi :\quad\T(\C P^2)\to\C P^2
\end{equation}
and the Riemann sphere $\C P^1$ as a typical fibre. We will endow this twistor
space with a one-parameter family of metric $\{g_{\s}\}_{|\s\in(0,\, \infty )}$
such that
this space is a nearly K\"ahler manifold for a special choice of the parameter $\s$.
Although the geometry of the coset space $\T(\C P^2)=\,$SU(3)/U(1)$\times$U(1)
is well-known, we describe it briefly by using local coordinates for fixing our
notation and further applications.

\smallskip

\noindent
{\bf Coset representatives}. Consider the principal bundle
\begin{equation}\label{2.6}
\mbox{SU}(3)\to\T(\C P^2)
\end{equation}
with the structure group U(1)$\times$U(1). Let $y^{\a}, \a =1,2$, be local
complex coordinates on $\C P^2$ and $\zeta$ a local complex coordinate on the
typical fibre $\C P^1$ of (\ref{2.5}). Then a local section of the fibration
(\ref{2.6}) (a representative element for the coset) is given by the 3$\times$3
matrix
\begin{equation}\label{2.7}
\hat V=\g^{-1}\begin{pmatrix}1&- T^{\+}\\T&W\end{pmatrix}
\begin{pmatrix}1&0\\0&h\end{pmatrix}\in
{\rm SU}(3)\ ,
\end{equation}
where
\begin{equation}\label{2.8}
T:=\begin{pmatrix}\yb^{\2}\\y^1\end{pmatrix} ,\quad
W:=\g\cdot{\bf 1}_2 - \frac{1}{\g +1}\, TT^{\+}\und
\g = (1+ T^{\+}T)^{\sfrac{1}{2}}=(1+ y^{\a}\yb^{\ab})^{\sfrac{1}{2}}
\end{equation}
obey
\begin{equation}\label{2.9}
 W^{\+}=W\ ,\quad WT=T\und W^2=\g^2\cdot{\bf 1}_2-TT^{\+}\ ,
\end{equation}
and therefore $V^{\+}V=V\,V^{\+}= {\bf 1}_3$. Here
\begin{equation}\label{2.10}
h=\frac{1}{(1+\zeta\bar\zeta)^{\frac{1}{2}}}\,
\begin{pmatrix}1&-\bar\zeta\\ \zeta & 1\end{pmatrix}
\in {\rm SU}(2)\cong S^3
\end{equation}
is a local section of the Hopf bundle $S^3\to S^2$ representing the coset $S^2\cong\,$SU(2)/U(1).

\smallskip

\noindent
{\bf Flat connection on $\T(\C P^2)$}.
Consider now a trivial complex vector bundle $\T(\C P^2)\times\C^4\to\T(\C P^2)$
endowed with a flat connection
\begin{equation}\label{2.11}
\hat\Acal = \hat V^{-1}\diff\hat V =:
 \begin{pmatrix}2b&-\sfrac{1}{2\La}\,\hat \t^{\+}\\
\sfrac{1}{2\La}\hat\t&\hat B\end{pmatrix} \ ,
\end{equation}
where
\begin{equation}\label{2.12}
 \hat\t = h^{\+}\t =\frac{1}{(1+\zeta\zeb)^{\frac{1}{2}}}
\begin{pmatrix}
\t^{\2}+\bar\zeta\t^1\\ \t^1-\zeta\t^{\2}
\end{pmatrix}
=:
\begin{pmatrix}
 \Th^{\2}\\ \Th^1
\end{pmatrix}\ ,\qquad
\hat\t^{\+}=\t^{\+}h=(\Th^2\ \Th^{\1})\ ,
\end{equation}
\begin{equation}\label{2.13}
 \hat B = h^{\+}B\,h +h^{\+}\diff\,h=:
\begin{pmatrix}
 \hat a_+& -\sfrac{1}{2R}\Th^{3}\\ \sfrac{1}{2R}\Th^{\3}&-\hat a_+
\end{pmatrix} -b\cdot{\bf 1}_2
\end{equation}
with
\begin{equation}\label{2.14}
\hat a_+ =\frac{1}{1+\zeta\zeb}\,\left\{ (1-\zeta\zeb)a_+ + \zeb b_+ -
\zeta\bar b_+ + \frac{1}{2}(\zeb\diff\zeta -\zeta\diff\zeb )\right\}\ ,
\end{equation}
\begin{equation}\label{2.15}
\Th^{3} =\frac{2R}{1+\zeta\zeb}\, \left(\diff\zeb  + \bb_+ + 2\zeb a_+ +
\zeb^2b_+\right)\ .
\end{equation}
Here $\t^1,\t^2, b, a_+$ and $b_+$ are defined by formulae~\cite{Popov2}
\begin{equation}\label{2.16}
 b=\frac{1}{4\g^2}(T^{\+}\diff T - \diff T^{\+} T)\ ,
\end{equation}
\begin{equation}\label{2.17}
\t=\frac{2\La}{\g^2}\, W\,\diff T= \begin{pmatrix}\t^{\2}\\
\t^1\end{pmatrix}=\frac{2\La}{\g}\,
\begin{pmatrix}\diff\yb^{\2}\\ \diff y^1\end{pmatrix}-
\frac{2\La}{\g^2(\g +1)}\, \begin{pmatrix}\yb^{\2}\\
y^1\end{pmatrix}(\yb^{\1}\diff y^{1} + y^2\diff\yb^{\2})\ ,
\end{equation}
\begin{equation}\label{2.18}
 \begin{pmatrix}a_+&-\bar b_+\\
b_+&-a_+\end{pmatrix}:=B + b\cdot{\bf 1}_2=
\frac{1}{\g^2}\,(W\diff W -T\,\diff T^{\+}- \frac{1}{2}\,\diff T^{\+}T -
\frac{1}{2}\,T^{\+}\diff T )+ b\cdot{\bf 1}_2\ .
\end{equation}
Note that $\t^1$ and $\t^2$ are local orthonormal basis of (1,0)-forms
on $\C P^2$. The real parameters $\La$ and $R$, inserted in the definition of
$\Th^1, \Th^2$ and $\Th^3$, respectively, characterize `sizes' of $\C P^2$
and $\C P^1\hra\T(\C P^2)=\,$SU(3)/U(1)$\times$U(1) from (\ref{2.5}).

\smallskip

\noindent
{\bf SU(3)-structures on $\T(\C P^2)$}. From flatness of the connection (\ref{2.11})
we obtain
\begin{equation}\label{2.19}
\diff b = -\frac{1}{8\La^2}\, (\Th^1\wedge\Th^{\1}-\Th^2\wedge\Th^{\2})\ ,
\quad \diff\ah_+ =  -\frac{1}{8\La^2}\,
(\Th^1\wedge\Th^{\1}+\Th^2\wedge\Th^{\2}-\frac{2}{\s}\,\Th^3\wedge\Th^{\3})
\end{equation}
with
\begin{equation}\label{2.20}
\s :=\frac{R^2}{\La^2}\ ,
\end{equation}
and the structure equations
\begin{equation}\label{2.21}
\diff\begin{pmatrix}\Theta^1\\ \Theta^2\\ \Theta^3\end{pmatrix}=
\begin{pmatrix}\hat a_++3b&0&0\\
0&\hat a_+-3b&0\\
0& 0& -2\hat a_+
\end{pmatrix}
\wedge
\begin{pmatrix}\Theta^1\\ \Theta^2\\ \Theta^3\end{pmatrix}
+ \frac{1}{2R}
\begin{pmatrix}\Theta^{\2}\wedge\Theta^{\3}\\ \Theta^{\3}\wedge\Theta^{\1}
\\ \s\Theta^{\1}\wedge\Theta^{\2}\end{pmatrix}\ ,
\end{equation}
where the first term defines $u(1)\oplus u(1)$ torsional connection
and the last term defines the Nijenhuis tensor (torsion) with components
$N^i_{\jb\kb}$ and their complex conjugate. Namely, we have
\begin{equation}\label{2.22}
 N^1_{\2\3}=N^2_{\3\1}=\frac{1}{2R}\und N^3_{\1\2}=\frac{R}{2\La^2}=\frac{\s}{2R}\ .
\end{equation}

We see that the parameters $\La$ and $R$ enter in all formulae but of
real importance is only their ratio (\ref{2.20}). For instance, consider the forms
$\Om$ and $\ome$ defined by (\ref{2.2}) and (\ref{2.4}). From (\ref{2.21})
it follows that
\begin{equation}\label{2.23}
\diff\ome=\frac{1}{2R}\, (2+\s )\,\mbox{Im}\Om\ ,\quad\diff\Om=\frac{1}{2R}\,
(\Th^2\wedge \Th^3\wedge\Th^{\2}\wedge\Th^{\3}+
\Th^3\wedge \Th^1\wedge\Th^{\3}\wedge\Th^{\1}+ \s\Th^1\wedge \Th^2\wedge
\Th^{\1}\wedge\Th^{\2})\ ,
\end{equation}
and we have a one-parameter\footnote{Overall scaling parameter is not essential,
$R$ can be fixed to some number.} family $(\ome , \Om )$ of SU(3)-structures on
SU(3)/U(1)$\times$U(1) such that for $\s=1$ this manifold is nearly K\"ahler
(see e.g.~\cite{CS, But, Xu} and references therein).

\bigskip

\section{SU(4)-instanton equations in eight dimensions}
\smallskip

\noindent
{\bf Instanton equations in $d>4$}. The concept of Yang-Mills instantons in four
dimensions can be generalized by considering first-order equations for gauge
potentials in spaces of dimension greater than four~\cite{CDFN}-\cite{IP93},
\cite{DT, Lewis}, \cite{Car}-\cite{Wolf}. Most of these equations naturally
appear in superstring theory and M-theory as the conditions for the survival
of at least one unbroken supersymmetry in low-energy effective field theory
in $d\le 4$ dimensions. Generically BPS-type first-order gauge equations in
higher dimensions can only be defined on manifolds with a $G$-structure. For
instance, on K\"ahler, Calabi-Yau and hyper-K\"ahler manifolds one can define
the Hermitian-Yang-Mills equations~\cite{Don, UY}. In $d=7$ one should consider
manifolds with a $G_2$-structure~\cite{DT, Car, Tian, DonSe}. In $d=8$ one
should consider a reduction of the holonomy group to Spin(7), SU(4), Sp(2)Sp(1)
or Sp(2) subgroups of SO(8)~\cite{DT, Lewis, Cap}, \cite{Car}-\cite{DonSe} and so on.
Some solutions of various type of the above-mentioned first-order gauge equations were
found e.g. in~\cite{group1, group2}. Studying the geometry of moduli spaces of
Yang-Mills instantons in $d>4$ is considered as an important
task~\cite{DT, Bau, Tian, DonSe}.

\smallskip

\noindent
{\bf $\Xi$-anti-self-duality}. BPS-type instanton equations in more than
four dimensions can be introduced as follows. Let $(X^d, g)$ be an oriented
Riemannian manifold of dimension $d$ and $\Xi$ a differential form of degree
$d-4$ on $X^d$. Consider a complex vector bundle $\Ecal$ over $X^d$ endowed
with a connection $\Acal$. The $\Xi$-anti-self-dual gauge equations are
defined~\cite{Tian} as the first-order equations,
\begin{equation}\label{3.1}
*\Fcal =-\Xi\wedge\Fcal\ ,
\end{equation}
on a connection $\Acal$ with the curvature $\Fcal =\diff\Acal +
\Acal\wedge\Acal$.\footnote{In four dimensions, (\ref{3.1}) is reduced
to $*\Fcal =-\Fcal$.}
Here $*$ is the Hodge star operator.

Differentiating (\ref{3.1}), we obtain
\begin{equation}\label{3.2}
\diff\!*\!\Fcal+\Acal\wedge*\Fcal -*\Fcal\wedge\Acal+*H\wedge\Fcal =0\ ,
\end{equation}
where the 3-form $H$ is defined by the formula
\begin{equation}\label{3.3}
H:=*\,\diff\Xi\ .
\end{equation}
Equations (\ref{3.2}) differ from the standard Yang-Mills equations
by the last term with a 3-form $H$ which can be identified with a totally
antisymmetric torsion. This torsion term naturally appears in string theory
(a lot of references can be found in~\cite{Popov2}).

\smallskip

\noindent
{\bf Spin(7)-instantons}. Let us consider a Riemannian manifold $(X^8, g)$
of dimension 8, and let $\widetilde\Xi$ be a 4-form which defines an almost
Spin(7)-structure\footnote{One can omit the word `almost' if $\widetilde\Xi$
is closed~\cite{Lewis, Tian, DonSe, Joy}. However, the case $\diff\widetilde\Xi\ne 0$
is also of interest and with an air of importance~\cite{Br, DonSe, Hay}.} on $X^8$.
A Spin(7)-instanton is defined as a connection $\widetilde\Acal$ on a complex vector
bundle $\widetilde\Ecal$ over $X^8$ such that its curvature $\widetilde\Fcal$ satisfies
the equations
\begin{equation}\label{3.4}
*\widetilde\Fcal =-\widetilde\Xi\wedge\widetilde\Fcal
\end{equation}
with an almost Spin(7)-structure $\widetilde\Xi$ on $X^8$.
For more details see e.g.~\cite{DT, Lewis, DonSe, Br, Hay}.

\smallskip

\noindent
{\bf SU(4)-instantons}. Suppose that an almost Spin(7)-manifold
$(X^8, g)$ allows an almost complex structure $\J$ and an SU(4)-structure,
i.e. $c_1(X^8)=0$. Then on $X^8$ there exists a non-degenerate (4,0)-form
$\Ow$ and an (1,1)-form $\ot$ such that
\begin{equation}\label{3.5}
\Ow=\Th^1\wedge\Th^2\wedge\Th^3\wedge\Th^4\and
\ot=\sfrac{\im}{2}\,(\Th^1\wedge\Th^{\1}+\Th^2\wedge\Th^{\2}+\Th^3\wedge\Th^{\3}
+\Th^4\wedge\Th^{\4})\ ,
\end{equation}
with
\begin{equation}\label{3.6}
\J\,\Th^A=\im\,\Th^A\for A=1,...,4\ ,
\end{equation}
i.e. $\Th^A$'s are (1,0)-forms with respect to $\J$.\footnote{For an integrable
almost complex structure $\J$ we obtain a Calabi-Yau 4-fold.} For such a case
the 4-form $\widetilde\Xi$ - an almost Spin(7)-structure - can be written as
\begin{equation}\label{3.7}
\widetilde\Xi = \sfrac{1}{2}\,\ot\wedge\ot - \mbox{Re}\,\Ow\ .
\end{equation}

The inclusion SU(4)$\ \subset\ $Spin(7) allows us to reduce the Spin(7)-instanton
equations (\ref{3.4}) with $\widetilde\Xi$ given in (\ref{3.7}) to SU(4)-instanton
equations~\cite{Tian}. Let $\widetilde\Ecal$ be a zero-degree bundle,
i.e. $c_1(\widetilde\Ecal )=0$ and therefore $\tr\widetilde\Fcal =0$. Then there
exists a Hermitian rank-$k$ vector bundle $\Ecal$ with a connection $\Acal$ such
that $\widetilde\Fcal =\Fcal - \sfrac{1}{k}(\tr\Fcal )\cdot{\bf 1}_k$, where
$\Fcal =\diff\Acal +\Acal\wedge\Acal$ is the curvature of $\Acal$. Note that
$\sfrac{\im}{2\pi}\,\tr\Fcal$ represents the first Chern class $c_1(\Ecal )$ in
$H^2(X^8,\R )$. Recall that on $X^8$ we are given a (4,0)-form $\Ow$ and its
complex conjugate $\bar\Ow$ induces an anti-linear involution $*^{}_{\Ow}:
\La^{0,2}(X^8)\to\La^{0,2}(X^8)$ so that one can introduce a self-dual part
\begin{equation}\label{3.8}
\Fcal^{0,2}_+:=\sfrac{1}{2}\,(\Fcal^{0,2}+*^{}_{\Ow}\,\Fcal^{0,2})
\end{equation}
of $\Fcal^{0,2}$. Assume now that $\tr\Fcal$ is harmonic and
$\tr\Fcal^{0,2}_+ =0$~\cite{Tian}. Then Spin(7)-instanton equations (\ref{3.4})
for $\widetilde\Acal = \Acal -\sfrac{1}{k}(\tr\Acal )\cdot{\bf 1}_k$ are equivalent
to the equations
\begin{equation}\label{3.9}
\im\,\ot\lrc\Fcal =\tilde\la\cdot{\bf 1}_k\und\Fcal^{0,2}_+=0\ ,
\end{equation}
where $\ot\lrc$ denotes a contraction of $\Fcal$ with a bivector dual to
$\ot$~\cite{CS} and $\tilde\la\in\R$ is related with $c_1(\Ecal )$.\footnote{For
$c_1(\Ecal )=0$ one has $\tilde\la =0$. In this case one can identify $\Ecal$
and $\widetilde\Ecal$.} Equations (\ref{3.9}) are called SU(4)-instanton
equations~\cite{DT, Tian}. In the basis of (1,0)-forms $\Th^A$ and (0,1)-forms
$\Th^{\bar A}$ they can be written as
\begin{equation}\label{3.10}
2\de^{A\bar A}\Fcal_{A\bar A}=\tilde\la\cdot{\bf 1}_k
\and
\Fcal_{\bar A\bar B}+
\sfrac{1}{2}\,\ve_{\bar A\bar B\bar C\bar D}\Fcal^{\bar C\bar D}=0\ ,
\end{equation}
where $\ve_{\bar A\bar B\bar C\bar D}$ is the totally skew-symmetric tensor.

\smallskip

\noindent
{\bf Hermitian-Yang-Mills equations}. Note that $\Fcal^{0,2}\ne 0$ for SU(4)-instantons
according to (\ref{3.9}) and (\ref{3.10}). However, on an almost complex manifold $X^8$
one can also introduce the Hermitian-Yang-Mills equations~\cite{Don, UY, Bryant},
\begin{equation}\label{3.11}
\im\,\ot\lrc\Fcal =\tilde\la\cdot{\bf 1}_k\und\Fcal^{0,2}=0\ ,
\end{equation}
which impose restrictions on a connection $\Acal$ on a complex vector bundle $\Ecal\to X^8$
stronger than equations (\ref{3.9}). Any solution $\Acal$ of (\ref{3.11}) solves
(\ref{3.9}) but the converse is not true. According to R.Bryant~\cite{Bryant},
any connection $\Acal$ on $\Ecal$ which satisfies (\ref{3.11}) defines a
pseudo-holomorphic structure $\bar\pa_{\Acal}$ on $\Ecal$. Solutions $\Acal$ of
(\ref{3.11}) are called the Hermitian-Yang-Mills connections. In the case of
integrable almost complex structure $\J$ the Hermitian-Yang-Mills
connections $\Acal$ define (poly)stable holomorphic bundles $\Ecal$~\cite{Don, UY}.

\smallskip

\noindent
{\bf Manifold $G/H\times\R^2$}. Let us consider the direct product
\begin{equation}\label{3.12}
G/H\times\R^2\ ,
\end{equation}
where $G/H$ is the homogeneous space SU(3)/U(1)$\times$U(1) described in section 2
or any other coset space with an SU(3)-structure, e.g. Sp(2)/Sp(1)$\times$U(1),
$S^6$ or $S^3\times S^3$. So, on $G/H$ we have an SU(3)-structure $(\ome , \Om )$,
a Hermitian metric $g$ and never integrable almost complex structure $\J$. On the
manifold (\ref{3.12}) we introduce an almost complex structure $\widetilde\J =
(\J, \mathfrak{j})$, where $\mathfrak{j}$ is the canonical (integrable) almost complex
structure on the space $\R^2$ with coordinates $x^7, x^8$. Namely, $\mathfrak{j}$ is
defined so that
\begin{equation}\label{3.13}
\Th^4:=\diff z^4 =\diff x^7 + \im\,\diff x^8
\end{equation}
is a (1,0)-form on $\R^2\cong\C$.

On $G/H\times\R^2$, we introduce forms
\begin{equation}\label{3.14}
\Ow = \Om\wedge\Th^4 = \Th^1\wedge\Th^2 \wedge\Th^3 \wedge\Th^4 \ ,
\end{equation}
\begin{equation}\label{3.15}
\ot = \ome +\sfrac{\im}{2}\,\Th^4\wedge\Th^{\4} =
\sfrac{\im}{2}\,(\Th^1\wedge\Th^{\1} + \Th^2 \wedge\Th^{\2} +
\Th^3\wedge\Th^{\3} + \Th^4\wedge\Th^{\4}) \ ,
\end{equation}
and the metric
\begin{equation}\label{3.16}
\widetilde g=g+\Th^4\Th^{\4}=\Th^1\Th^{\1}+\Th^2\Th^{\2}+\Th^3\Th^{\3}+
\Th^4\Th^{\4}\ .
\end{equation}
Thus, $G/H\times\R^2$ is an 8-dimensional Riemannian manifold with an
SU(4)-structure.\footnote{In fact, the torsional connection on $G/H\times\R^2$
has holonomy contained in SU(3) due to trivial holonomy along the subspace $\R^2$.}
Hence, on the manifold (\ref{3.12}) one can introduce the 4-form (\ref{3.7})
and the Spin(7)-instanton equations (\ref{3.4}). These equations can also be
rewritten in the form (\ref{3.9}) of SU(4)-instanton equations.

\bigskip

\section{Vortex equations associated with SU(3)/U(1)$\times$U(1)}

\noindent
In this section we will consider Yang-Mills theory with SU(3)-equivariant
gauge fields on the manifold $X^8=G/H\times\R^2$ with $G{=}$SU(3) and
$H{=}$U(1)$\times$U(1). The group $G$ acts trivially on $\R^2\cong\C$
and in the standard way by isometries on $G/H$.

\smallskip

\noindent
{\bf Invariant connection}. Let $\Ecal\to X^8$ be an SU(3)-equivariant complex
vector bundle of rank $k$ over $X^8$ and $\Acal$ a $u(k)$-valued local form
of SU(3)-equivariant\footnote{This means a generalized SU(3)-invariance, i.e.
invariance under SU(3)-isometries up to gauge transformations~\cite{FMT, Kub, KZ, G4}.
For transition functions one considers a compensating change of a trivialization.}
connection on $\Ecal$. Such a connection $\Acal$ is given by naturally
extending the flat connection $\hat\Acal$ on the bundle $\hat\Ecal\to G/H$
from (\ref{2.11})~\cite{LPS4}. In the simplest case of a quiver bundle $E^{0,1}$
 associated with the fundamental representation $\C^3$ of SU(3) we get
\begin{equation}\label{4.1}
\Acal = \begin{pmatrix}A^1\otimes 1 + {\bf 1}_{k_1}\otimes 2b&
-\sfrac{1}{2\La}\,\p^{\+}_2\otimes\Th^2&-\sfrac{1}{2\La}\,\p_1\otimes\Th^{\1}\\[6pt]
\sfrac{1}{2\La}\,\p_2\otimes\Th^{\2}&A^2\otimes 1 + {\bf 1}_{k_2}\otimes (\hat a_+-b)&
-\sfrac{1}{2R}\,\p^{\+}_3\otimes\Th^3\\[6pt]
\sfrac{1}{2\La}\,\p^{\+}_1\otimes\Th^1&\sfrac{1}{2R}\,\p_3\otimes\Th^{\3}&
A^3\otimes 1 - {\bf 1}_{k_3}\otimes (\hat a_+ +b)
\end{pmatrix} \ ,
\end{equation}
where $A^1, A^2$ and $A^3$ are $u(k_1)$-, $u(k_2)$- and $u(k_3)$-valued gauge
potentials on complex vector bundles $E_1, E_2$ and $E_3$ over $\R^2$ with ranks
$k_1, k_2$ and $k_3$, respectively, such that $k_1+k_2+k_3=k=\,$rank$\,\Ecal$.
The bi-fundamental scalar fields $\phi_1\in Hom (E_3, E_1), \phi_2\in Hom (E_1, E_2)$
and $\phi_3\in Hom (E_2, E_3)$ can be identified with sections (Higgs fields) of
the bundles $E_1\otimes E_3^{\vee}$, $E_2\otimes E_1^{\vee}$ and
$E_3\otimes E_2^{\vee}$, respectively. Note that fields $A^i$ and $\p_i$ depend
only on coordinates $x^7, x^8$ of $\R^2$ and $\p_i^\+$ is a Hermitian conjugate
of $\p_i, i=1,2,3$. Ansatz for the invariant connection $\Acal$ associated with
arbitrary representation $C^{p,q}$ of SU(3) can be written down by using the
formulae from~\cite{LPS4}. However, in this short paper we
intentionally will consider only the simplest case for exemplifying non-typical
properties of new quiver vortex equations.

\smallskip

\noindent
{\bf $C^{0,1}$-quiver bundle}. The fundamental representation $\C^3$ ($\sim C^{0,1}$)
of SU(3) decomposes as $\C^3 = \C\oplus\C\oplus\C$ after the restriction to
U(1)$\times$U(1). To this decomposition there corresponds a
U(1)$\times$U(1)-equivariant bundle
\begin{equation}\label{4.2}
E^{0,1}=E_1\otimes\C\oplus E_2\otimes\C\oplus E_3\otimes\C
\end{equation}
over $\R^2$ along with appropriate bundle morphisms $\p_i$ between $E_i$ given by the
quiver diagram
\bigskip
\begin{equation}\label{4.3}
\begin{picture}(50,40)
\put(0.0,0.0){\makebox(0,0)[c]{$E_1\,\bullet$}}
\put(50.0,0.1){\vector(-1,0){36}}
\put(66.0,0.0){\makebox(0,0)[c]{$\bullet\,E_3$}}
\put(32.0,38.0){\makebox(0,0)[c]{$E_2$}}
\put(32.0,30.0){\makebox(0,0)[c]{$\bullet$}}
\put(7.0,18.0){\makebox(0,0)[c]{$\p_2$}}
\put(55.0,18.0){\makebox(0,0)[c]{$\p_3$}}
\put(34.0,5.0){\makebox(0,0)[c]{$\p_1$}}
\put(7.0,5.0){\vector(1,1){20}}
\put(37.0,25.0){\vector(1,-1){20}}
\end{picture}
\end{equation}
\bigskip
We also introduce the homomorphisms
\begin{equation}\label{4.4}
r_1:=\p_1-\p_2^\+\p_3^\+\ ,\quad r_2:=\p_2-\p_3^\+\p_1^\+\und
r_3:=\p_3-\p_1^\+\p_2^\+
\end{equation}
corresponding to the quiver relations.

In (\ref{4.2}), factors $\C$ denote the trivial U(1)$\times$U(1)-equivariant
complex line bundles over $\R^2$ arising from the decomposition
$\C^3=\C\oplus\C\oplus\C$. Later we will see that in general  the diagram
(\ref{4.3}) is not commutative due to non-integrability of the Dolbeault
operator on vector bundles over the coset space SU(3)/U(1)$\times$U(1) with
non-integrable almost complex structure. This will result in the proportionality
of $\bar\pa_A\p_i$ to\footnote{Here, $\bar\pa_A$ is the Dolbeault operator
on the vector bundle $E^{0,1}$ over $\R^2$.} the polynomials $r_i$ from
(\ref{4.4}) corresponding to the quiver relations.

\smallskip

\noindent
{\bf Invariant field strength tensor.} For the curvature\footnote{By $i,j$ we
are numbering $k_i\times k_j$ blocks in $\Fcal$.} $\Fcal =\diff\Acal
+\Acal\wedge\Acal=(\Fcal^{ij})$ of the invariant connection $\Acal$ given by
(\ref{4.1}) we obtain
\begin{eqnarray}
\Fcal^{11}&=&F^{1}-\sfrac{1}{4\La^2}\,({\bf 1}_{k_1}-\p_1\p_1^\+)
\Th^1\wedge\Th^{\1}+\sfrac{1}{4\La^2}\,({\bf 1}_{k_1}-\p_2^\+\p_2)
\Th^2\wedge\Th^{\2}\ ,\label{4.5}\\[4pt]
\Fcal^{22}&=&F^{2}-\sfrac{1}{4\La^2}\,({\bf 1}_{k_2}-\p_2\p_2^\+)
\Th^2\wedge\Th^{\2}+\sfrac{1}{4R^2}\,({\bf 1}_{k_2}-\p_3^\+\p_3)
\Th^3\wedge\Th^{\3}\ ,\label{4.6}\\[4pt]
\Fcal^{33}&=&F^{3}+ \sfrac{1}{4\La^2}\,({\bf 1}_{k_3}-\p_1^\+\p_1)
\Th^1\wedge\Th^{\1}-\sfrac{1}{4R^2}\,({\bf 1}_{k_3}-\p_3\p_3^\+)
\Th^3\wedge\Th^{\3}\ ,\label{4.7}\\[4pt]
\Fcal^{13}&=& -\sfrac{1}{2\La}\,(\diff\phi_{1}+A^{1}\phi_{1}-
\phi_{1}A^{3} )\wedge\Th^{\1}-\sfrac{1}{4\La R}\,(\p_1-\p_2^\+\p_3^\+)
\Th^2\wedge\Th^{3}\ ,\label{4.8}\\[4pt]
\Fcal^{21}&=& \sfrac{1}{2\La}\,(\diff\phi_2 + A^{2}\phi_{2}-
\phi_{2}A^{1} )\wedge\Th^{\2}+\sfrac{1}{4\La R}\,(\p_2-\p_3^\+\p_1^\+)
\Th^3\wedge\Th^{1}\ ,\label{4.9}\\[4pt]
\Fcal^{32}&=& \sfrac{1}{2R}\,(\diff\phi_3 + A^{3}\phi_{3}-
\phi_{3}A^{2})\wedge\Th^{\3}+\sfrac{1}{4\La^2}\,(\p_3-\p_1^\+\p_2^\+)
\Th^1\wedge\Th^{2}\ ,\label{4.10}
\end{eqnarray}
plus their Hermitian conjugates $\Fcal^{ji}=-(\Fcal^{ij})^{\+}$ for $i\ne j$,
$i,j,...=1,2,3$. In deriving (\ref{4.5})-(\ref{4.10}) we used various relations
following from the flatness of the connection (\ref{2.11}). In (\ref{4.5})-(\ref{4.10})
$F^i=\diff A^i + A^i\wedge A^i$ is the curvature of a connection $A^i$ on
the complex vector bundle $E_i$.

\smallskip

\noindent
{\bf Quiver vortex equations.}  Let us substitute the field strength matrix elements
(\ref{4.5})-(\ref{4.10}) and their Hermitian conjugates into the SU(4)-instanton
equations (\ref{3.10}). After this substitution we get non-Abelian coupled vortex
equations
\begin{eqnarray}
\pa_{\bar z}\phi_{1}+A^{1}_{\bar z}\phi_{1}-\phi_{1}A^{3}_{\bar z}=
\sfrac{1}{2 R}\,(\p_1-\p_2^\+\p_3^\+)\ ,\label{4.11}\\[4pt]
\pa_{\bar z}\phi_{2}+A^{2}_{\bar z}\phi_{2}-\phi_{2}A^{1}_{\bar z}=
\sfrac{1}{2 R}\,(\p_2-\p_3^\+\p_1^\+)\ ,\label{4.12}\\[4pt]
\pa_{\bar z}\phi_{3}+A^{3}_{\bar z}\phi_{3}-\phi_{3}A^{2}_{\bar z}=
\sfrac{\s}{2 R}\,(\p_3-\p_1^\+\p_2^\+)\ ,\label{4.13}\\[4pt]
F^{1}_{z\bar z}=\sfrac{1}{4R^2}\,(\la\cdot{\bf 1}_{k_1}-\s\p_1\p_1^\+
+\s\p_2^\+\p_2)\ ,\label{4.14}\\[4pt]
F^{2}_{z\bar z}=\sfrac{1}{4R^2}\,((\la{+}\s{-}1){\bf 1}_{k_2} -\s\p_2\p_2^\+
+\p_3^\+\p_3)\ ,\label{4.15}\\[4pt]
F^{3}_{z\bar z}=\sfrac{1}{4R^2}\,((\la{-}\s{+}1){\bf 1}_{k_3} -\p_3\p_3^\+
+\s\p_1^\+\p_1)\ ,\label{4.16}
\end{eqnarray}
where $\la := 2\tilde\la R^2$ and $z:=z^4=x^7+\im\,x^8\ , \bar z:=\bar z^{\4}$.
Recall that $\s = R^2/\La^2$. Note that one can consider these equations not
only on $\R^2$ but also on $\R\times S^1$ and on a complex torus $T^2=S^1\times S^1$.

Let us denote by $\bar\pa_A$ the Dolbeault operator acting on $\p_i$ in
(\ref{4.11})-(\ref{4.13}). From these equations we see that the Higgs fields
$\p_i$, defining homomorphisms of the bundles $E_i\to\R^2$, are {\it not}
holomorphic. In fact, $\bar\pa_A\p_i$ is proportional to the quiver relation
$r_i$ defined in (\ref{4.4}),
\begin{equation}\label{4.17}
\bar\pa_A\p_i = \vk\, r_i\ ,
\end{equation}
where $\vk =\diff\zb/2R$ for $i=1,2$ and $\vk =\s\diff\zb/2R$ for $i=3$.
This reflects the fact that the almost complex structure $\widetilde\J$ on
$\R^2\times\,$SU(3)/U(1)$\times$U(1) is not integrable, the bundle $\Ecal$
is not (pseudo)holomorphic and therefore $\Fcal^{0,2}$ must not vanish.
Of course, on the SU(3)-equivariant connection $\Acal$ on $\Ecal$ one can impose
the Hermitian-Yang-Mills equations (\ref{3.11}) which are stronger than (\ref{3.9}).
Then one shall get the standard quiver vortex equations
\begin{equation}\label{4.18}
\bar\pa_A\p_i = 0\und r_i=0
\end{equation}
together with (\ref{4.14})-(\ref{4.16}). However, there are solutions to
(\ref{4.14})-(\ref{4.17}) which are not reduced to solutions of
(\ref{4.14})-(\ref{4.16}) and (\ref{4.18}).

\smallskip

\noindent
{\bf Particular cases.} Let us put $\s =1$ when the SU(3)-structure on
SU(3)/U(1)$\times$U(1) becomes nearly K\"ahler. Then the quiver vortex
equations (\ref{4.11})-(\ref{4.16}) become more `symmetric':
\begin{eqnarray}
D_{\bar z}\phi_{i}=\sfrac{1}{2 R}\,r_i\ \ \mbox{with}\ \
\diff\zb\,D_{\zb}\p_i:=\bar\pa_A\p_i\nonumber\\[4pt]
F^{1}_{z\bar z}=\sfrac{1}{4R^2}\,(\la\cdot{\bf 1}_{k_1} -
\p_1\p_1^\+ +\p_2^\+\p_2)\ ,\nonumber\\[4pt]
F^{2}_{z\bar z}=\sfrac{1}{4R^2}\,(\la\cdot{\bf 1}_{k_2}
-\p_2\p_2^\+ +\p_3^\+\p_3)\ ,\label{4.19}\\[4pt]
F^{3}_{z\bar z}=\sfrac{1}{4R^2}\,(\la\cdot{\bf 1}_{k_3} -\p_3\p_3^\+
+\p_1^\+\p_1)\ .\nonumber
\end{eqnarray}
Choosing in these equations $k_1=k_2=k_3=:m$ and $\p_1=\p_2=\p_3=:\p$,
we obtain the vortex equations
\begin{equation}\label{4.20}
\bar\pa_{\zb}\p + [A_{\zb}, \p ] = \sfrac{1}{2R}\,(\p -\p^\+\p^\+)
\und
F_{z\bar z}=\sfrac{1}{4R^2}\,(\la\cdot{\bf 1}_{m} -[\p , \p^\+ ])\ ,
\end{equation}
where $F=\diff A + A\wedge A$ and $A:=A_1=A_2=A_3$. Here, the equality
of $A^i$'s follows from the equations for $F^i$, $i=1,2,3$.

Finally, for $\la =0$ and $[\p , \p^\+ ]=0$, we obtain $F=0=A$ and
(\ref{4.20}) is reduced to the equation
\begin{equation}\label{4.21}
\bar\pa_{\zb}\p = \sfrac{1}{2R}\,(\p -\p^\+\p^\+)
\end{equation}
on a matrix-valued scalar field. This equation obviously has nontrivial
solutions that proves the existence of nonholomorphic solutions to the
quiver vortex equations (\ref{4.11})-(\ref{4.16}).

\smallskip

\noindent
{\bf Vortices in $\Ncal=4$ super-Yang-Mills theory.}
Consider $\Ncal =1$ SYM theory with the gauge group U($k$) on ten-dimensional
flat space with the metric
\begin{equation}\label{4.22}
g_{10} = - \diff t^2 +\diff s^2 + \diff z\,\diff\zb + \Th^1\Th^{\1}+
\Th^2\Th^{\2}+\Th^3\Th^{\3}\ ,
\end{equation}
where $t, s, z$ and $\zb$ are coordinates on Minkowski space $\R^{3,1}$
and $\{\Th^i, \Th^{\ib}\}$ is the orthonormal complex frame on flat complex
`internal' space $\R^6\cong\C^3$ (or complex torus $T^6$). Considering
trivial dimensional reduction (invariance under translations) of $\Ncal =1$
SYM theory from $d=10$ to $d=4$, we obtain the standard $\Ncal =4$ SYM theory in
Minkowski space. Let us now reduce the bosonic sector of this theory to the
space $\R^2\cong\C$ with coordinates $z,\zb$ and assume that $\Acal_t=\Acal_s=0.$
Then after turning on a Fayet-Iliopoulos parameter $\la:=\sfrac12\tilde\la$ in
the Lagrangian, we see that the first-order BPS equations are dimensional reduction
of the SU(4)-instanton equations (\ref{3.10}) to $d=2$. Namely, we get the vortex
equations
\begin{equation}\label{4.23}
F_{z\bar z}=\la\cdot{\bf 1}_{k} -[\p_1, \p_1^\+]-[\p_2,\p_2^\+]-[\p_3,\p_3^\+ ]\ ,
\end{equation}
\begin{equation}\label{4.24}
\bar\pa_{\zb}\p_i + [A_{\zb}, \p_i] = \sfrac{1}{2}\,\ve_{ijk}[\p_j^\+,\p_k^\+]\ ,
\end{equation}
where $F_{z\bar z}=\pa_z A_{\zb} - \pa_{\zb}A_z + [A_z, A_{\zb}]\in u(k)$
and $\p_i$ are $k\times k$ complex matrices.\footnote{Of course, one can
assume that $\tr F=0$ and consider equations (\ref{4.23}), (\ref{4.24})
with $\la =0$.} Note that $\Phi_i:=\,$Re\,$\p_i$ and $\Phi_{i+3}:=\,$Im\,$\p_i$
take values in $u(k)$ and belong to the fundamental
representation of SO(6)$\cong$SU(4). For transition to the SU(4) notation
$\{\Phi_i, \Phi_{i+3}\}\to \p_{AB}= -\p_{BA}$ one should simply convert
$\Phi_i$, $\Phi_{i+3}$ with  the chiral part $\s^i_{AB}$, $\s^{i+3}_{AB}$
of $\g$-matrices in $d{=}6$: $\p_{AB}:= \Phi_i\s^i_{AB}{+}\Phi_{i+3}\s^{i+3}_{AB}$.

Non-Abelian BPS vortices can and do occur in $\Ncal =2$ supersymmetric gauge
theories (see e.g.~\cite{group3} and references therein). Contrary to the
$\Ncal\le 2$ case, vortices in $\Ncal =4$ SYM theory are described by
non-holomorphic  Higgs fields subject to the equations
(\ref{4.23}), (\ref{4.24}). These non-Abelian vortex equations allow further
(algebraic) reductions. For instance, if we put $\la =0$ and assume
$[\p_i,\p_i^\+]=0$, then from (\ref{4.23}) it follows
that $F=0=A$ and (\ref{4.24}) are reduced to the equations
\begin{equation}\label{4.25}
\bar\pa_{\zb}\p_i = \sfrac{1}{2}\,\ve_{ijk}[\p_j^\+,\p_k^\+]\ ,
\end{equation}
which generalize the celebrated Nahm equations and coincide with them for
anti-hermitian $\phi_i$.

\bigskip

\section{Conclusions}

\noindent
In this paper we have considered the Spin(7)-instanton equations on the almost
complex manifold $\R^2\times\,$SU(3)/U(1)$\times$U(1) with a family of
SU(4)-structures parametrized by $\s\in (0, +\infty )$. On such a manifold
Spin(7)-instanton equations are reduced to the SU(4)-instanton equations. We
considered an SU(3)-equivariant connection $\Acal$ on a complex vector bundle
$\Ecal$ over $\R^2\times\,$SU(3)/U(1)$\times$U(1) corresponding to a simplest
quiver associated with the coset space SU(3)/U(1)$\times$U(1). It has been shown
that for symmetric connections $\Acal$ the SU(4)-instanton equations on
$\R^2\times\,$SU(3)/U(1)$\times$U(1) are reduced to quiver vortex equations
on $\R^2$ with non-holomorphic coupled Higgs fields $\p_i , \ i=1,2,3$. It was
shown that `deviation' from the holomorphicity is given by the quiver relations
$r_i$, i.e. $\bar\pa_{A}\p_i = \vk\,r_i$. From these equations it follows that
the quiver diagram is not commutative until $\bar\pa_{A}\p_i = 0$ and $r_i=0$
separately. This happens when the connection $\Acal$ on $\Ecal$ satisfies the
more restrictive Hermitian-Yang-Mills equations. In general, the same is true
for any quiver with relations associated with the space SU(3)/U(1)$\times$U(1).
It was shown that the obtained quiver vortex equations can further be reduced
to a matrix kink-type equations (cf.~\cite{group2, Popov3}).

We have introduced BPS
vortex equations in $\Ncal =4$ supersymmetric Yang-Mills theory and shown that
they are also related with the SU(4)-instanton equations on flat eight-dimensional
space. It would be interesting to consider the above correspondence between
non-Abelian vortices on $\R^2$ and symmetric instantons on $\R^2\times X^6$ for
other 6-dimensional spaces with an SU(3)-structure, e.g. Sp(2)/Sp(1)$\times$U(1),
$S^6$ and others, as well as to construct exact solutions to the obtained vortex
equations.

\bigskip

\section*{Acknowledgments}

\noindent
This work was partially supported by the Deutsche Forschungsgemeinschaft
(grant 436 RUS 113/995) and the Russian Foundation for Basic Research
(grants 08-01-00014-a and 09-02-91347).

\bigskip

\end{document}